# Spherically symmetric event horizons and trapped surfaces developing from innociuous data

Ulrich Alfes and Henning Müller zum Hagen

Universität der Bundeswehr, Fachbereich Maschinenbau, D-22039 Hamburg

e-mail: m_alf@unibw-hamburg.de

**Abstract.** In this paper we show the existence of a large class of spherically symmetric data $d$ (on a spacelike hypersurface $S$), from which a perfect fluid spacetime (surrounded by vacuum) develops. This spacetime contains an event horizon (with trapped surfaces behind it). The data $d$ are regular and *innociuous*, i.e. the data–surface $S$ does not contain any point of the horizon or of the trapped surface area. The occurence of the horizon (and trapped surfaces) is stable under small spherically symmetric variations of the data $d$.

We give auxiliary data on an auxiliary hypersurface $H$ and also on the star boundary; then we solve Einstein's equations for perfect fluid in the future and past of $H$. Our solution induces the above mentioned data $d$ on some chosen spacelike hypersurface $S$ in the past of $H$. By construction $H$ turns out to be the matter part of the horizon, once we attach a vacuum to our matter spacetime. Obviously, from these data $d$ on $S$ it develops (into the future) the event horizon $H$.

We solve the constraint equations for the auxiliary data posed on the null–surface $H$. This reduces the choice of these data to the choice of the density $\rho$ and $R := [\text{the orbit curvature}]^{-1/2}$. Our data fulfil positivity of $\rho$, $2m/R = 1$ (at the star boundary) and other properties. This is archieved by an algorithm, which for given $\rho$ yields $R$ (from an input parameter function $h \in C^1([0,1], ]0, -\infty[)$).

## 1. Introduction

Let us confine ourselves to spherically symmetric spacetimes. In many interesting papers (e.g. [1],[2]) there was shown that sufficient mass–concentration induces the forming of trapped surfaces: certain conditions have been posed on *data $d$* given on some spacelike hypersurface $S$; it was shown that these conditions imply the existence of *trapped surfaces* on $S$.

One would like to extend these results into three directions

(i) The data on $S$ should be *innociuous*, i.e. the trapped surface should not occur on $S$, but rather in the *future* of $S$;

(ii) One would like to show the forming of an *event horizon* (a global object) along with trapped surfaces (a local object); (see [3], where null–surfaces are used instead of the spacelike surface $S$)

(iii) One would like to show the *existence* of a (unique) spacetime, which develops from the data $d$. In [2] (and also in [3], [4]), however the existence of such a spacetime has been *assumed*.

In the present paper we pose auxiliary data $d_H$ on an auxiliary hypersurface† $H$, (which will turn out to be part of the event horizon); furthermore, we pose data on the star boundary. Then we show the local (in time) existence of a (unique) solution‡ $V_1$ assuming the above data; $V_1$ is a perfect fluid spacetime containing also a part, which lies in the past of $H$. Now, in the past of $H$ we choose a spacelike hypersurface $S$; our solution $V_1$ induces data $d$ on $S$. From these data $d$ on $S$ evolves – into the future of $S$ – a unique solution; this solution obviously is our above solution $V_1$, which contains the hypersurface $H$ in the future of $S$. The null–extension of $H$ into the vaccuum, which we attach uniquely to $V_1$, is (by construction) an *event horizon*.

The above construction means the following: we have constructed data $d$ (on some spacelike hypersurface $S$), from which an *event horizon* (with trapped surfaces in its future) evolves. The occurence of the event horizon is *stable* (under spherically symmetric variations of the data $d$). The *properties (i),(ii),(iii)* are fulfilled. However, the mass–concentration will not be controlled on $S$ but on $H$, on which the auxiliary data $d_H$ are given. The mass–concentration is controlled as follows (see Remark 13 c), d)):

$$\frac{2m}{R} < 1 \text{ on } H \setminus B, \qquad \frac{2m}{R} = 1 \text{ on } H \cap B, \tag{1}$$

where $B$ is the star boundary and $m$ denotes Misner's mass§ scalar field (see [5]–[7]) and $R(P) :=$ [curvature of the orbit (of spherical symmetry) at the point $P$]$^{-1/2}$. The condition (1) is *sharp*: if we would replace (1) by $\frac{2m}{R} > 1$ on $H$, then $H$ would not be a horizon (see also Remark 13 c), d)).

It may be possible to extend the results of [2] such that (i) is fulfilled; Moreover, (iii) may also be satisfied in a neighbourhood of $S$ (using existence theorems for local solutions). However, in order to fulfil the global property (ii), one would need some *global*‖ *existence* theorem on solutions of Einstein's equations for perfect *fluid*. It will be difficult to obtain such global existence theorem, since the fluid can evolve shell–crossings or shocks.

One can avoid such difficulties, if one couples Einstein's equations not to a perfect fluid but to a scalar field, which has been done in [8]. A global existence theorem was proved in [8], and the properties (i),(ii),(iii) are also fulfilled.

The first author is writing his doctorial thesis, which includes the material of this paper as well as some numerical investigations. For details we refer to this thesis.

## 2. General assumptions, definitions, and coordinates

We shall *assume* that our spacetime is *spherically symmetric* (with one timelike centreline) and that there is given a smooth *equation of state* $\hat{p}(\rho)$ with density $\rho$ and pressure $p$, i.e.

$$p(x^\mu) = \hat{p}(\rho(x^\mu))$$

---

† see Remark 9 a).
‡ Actually, we shall construct this solution $V_1$ in two steps; (see Figure 1, where $V_1$ is in fact $V_F \cup V_H$).
§ This mass has been introduced by Tolman [9].
‖ In this paper we avoid using global solutions by choosing our (auxiliary) data–surface $H$ to be a *null*–surface.

(indices $\mu, \nu \ldots$ run from 1 to 4) fulfilling

$$0 < \frac{\mathrm{d}\hat{p}}{\mathrm{d}\rho} < 1, \ \hat{p}(\rho) > 0 \text{ for } \rho > \rho_{\mathrm{b}} \text{ and } \hat{p}(\rho_{\mathrm{b}}) = 0, \qquad (2)$$

where $\rho_{\mathrm{b}} > 0$ is some constant†.

We consider *Einstein's equations for perfect fluid*

$$\frac{1}{8\pi}G^{\mu\nu} = T^{\mu\nu} := (\rho + \hat{p}(\rho))u^{\mu}u^{\nu} + \hat{p}(\rho)g^{\mu\nu}, \ u^{\mu}u_{\mu} = -1$$

where $u^{\mu}$ denotes the fluid vector and $g^{\mu\nu}$ the metric field. We *define* $D_t := u^{\mu}\partial_{x^{\mu}}$ ($\partial_{x^{\mu}} := \frac{\partial}{\partial x^{\mu}}$) and the unit vector field $D_{\bar{r}}$ as being orthogonal to $D_t$ and to the orbits (of symmetry). Later, we shall consider a given $(r,t)$-coordinate plane $((r,t)$ of Definition 2). For $r = 0$ we *define* $D_{\bar{r}}$ as its continuous extension, provided we deal with spacetimes endowed with a regular centreline at $r = 0$ (e.g. in Lemma 5). Moreover, we *define* the *scalar fields*

$$R := 1/\sqrt{[\text{curvature of the orbits}]}, \quad U := D_t R \text{ and } \Gamma := D_{\bar{r}}R \qquad (3)$$

Furthermore, we *define* the following auxiliary functions

$$\hat{\phi}(\rho) := -\int_{\rho_{\mathrm{b}}}^{\rho} \frac{\mathrm{d}\hat{p}(x)}{\mathrm{d}x}[(\hat{p}(x) + x)]^{-1}\mathrm{d}x, \qquad \hat{n}(\rho) := n_{\mathrm{b}}\exp\left[\int_{\rho_{\mathrm{b}}}^{\rho} \frac{\mathrm{d}x}{(\hat{p}(x) + x)}\right], \qquad (4)$$

where $\rho_{\mathrm{b}} > 0$ (cf (2)), $n_{\mathrm{b}} > 0$ are some constants.

*Lemma 1.* a) Given $\rho_{\mathrm{H}}$ and $R_{\mathrm{H}}$ as real-valued functions of $r \in [0, r_{\mathrm{b}}]$. We *define*

$$B_{\mathrm{H}}(r) := \int_0^r \mathrm{e}^{-\hat{\phi}(\rho_{\mathrm{H}}(x))}\mathrm{d}x, \qquad A_{\mathrm{H}}(r) := \int_0^r 4\pi\,\hat{n}(\rho_{\mathrm{H}}(x)) \cdot R_{\mathrm{H}}^2(x)\,\mathrm{d}x. \qquad (5)$$

We *assume* that $(\mathrm{d}s^2, u^{\mu}\partial_{x^{\mu}}, \rho)$ is a spherically symmetric perfect fluid spacetime. Then there exists a coordinate system such that

$$\mathrm{d}s^2 = \mathrm{e}^{\lambda}\mathrm{d}r^2 + R^2\left(\mathrm{d}\vartheta^2 + (\sin^2\vartheta)\,\mathrm{d}\varphi^2\right) - \mathrm{e}^{2\phi}\left(\mathrm{d}t + B_{\mathrm{H}}'(r)\,\mathrm{d}r\right)^2 \qquad (6)$$

with $\lambda$, $R$, $\phi$ depending on $(r,t)$ (a dash denotes $\frac{\mathrm{d}}{\mathrm{d}r}$) and with

$$u^{\mu}\partial_{x^{\mu}} = \mathrm{e}^{-\phi}\partial_t, \qquad \text{(i.e. } co\text{-}moving \text{ coordinates)} \qquad (7)$$

for $r \in [0, r_{\mathrm{b}}], t \in [-T, T]$, where $r = 0$ (and $r = r_{\mathrm{b}}$, respectively) denotes the centreline (and the star boundary, respectively). It follows that the scalar field

$$A(r,t) := \int_0^r 4\pi\,\hat{n}(\rho(x,t)) \cdot R^2(x,t) \cdot \mathrm{e}^{\frac{1}{2}\lambda(x,t)}\mathrm{d}x \quad \text{depends on } r \text{ only}. \qquad (8)$$

b) *In addition we assume*

$$\rho = \rho_{\mathrm{H}} \text{ and } R = R_{\mathrm{H}} \text{ on } H := \{(r,t) \mid t = 0, r \in [0, r_{\mathrm{b}}]\}. \qquad (9)$$

Now, we can (and shall) *fix the coordinates* of (6,7) by

$$\phi(r_{\mathrm{b}}, t) = \hat{\phi}(\rho_{\mathrm{H}}(r_{\mathrm{b}})) \text{ (for } t \in [-T, T]) \qquad A(r, 0) = A_{\mathrm{H}}(r) \text{ (for } r \in [0, r_{\mathrm{b}}]) \qquad (10)$$

† We use the more convenient notations $\hat{p}(\rho)$ and $\frac{\mathrm{d}\hat{p}}{\mathrm{d}\rho}$ instead of $\hat{p} \circ \rho$ and $\hat{p}' \circ \rho$.

The latter is equivalent to $\lambda = 0$ on $H$. It follows that $H$ is a null–surface. Furthermore, $\phi$ and $\lambda$ of (6) are calculated as

$$\phi(r,t) = \hat{\phi}(\rho(r,t)), \qquad e^{\lambda/2} = \Gamma^{-1}(\partial_r - B'_H \partial_t) R. \tag{11}$$

Moreover (due to (8,10)), it holds

$$A(r,t) = A_H(r). \tag{12}$$

*Definition 2.* We denote the coordinates fixed by (6,7,10) as $(r,t)$-*coordinates*.

*Remark on Lemma 1.* Our $(r,t)$–coordinates are closely related to Misner's coordinates (see (13,14)). — Our $(r,t)$–coordinates are co–moving and spherically symmetric. It holds $t = 0$ on $H$, which is convenient, when we shall give $(\rho_H, R_H)$ as '*data*' on the surface $H$. These 'data' $(\rho_H, R_H)$ have been used for fixing the $(r,t)$–coordinates (see (10)). Incidentally, any quantity carrying an index $H$ can be calculated from these functions $(\rho_H, R_H)$, see e.g. (43). These quantities depend on the variable $r \in [0, r_b]$ only. — Note, that $\{(r,t) \mid t = t_0\}$ is in general *not* a null–surface, provided $t_0 \neq 0$.

*Proof of Lemma 1.* Let us denote by $(\bar{r}, \bar{t})$ the following coordinates used in [5]–[7]:

$$\begin{aligned} ds^2 &= g_{\bar{r}\bar{r}}(\bar{r}, \bar{t}) d\bar{r}^2 + R^2(\bar{r}, \bar{t})(d\vartheta^2 + (\sin^2\vartheta) d\varphi^2) + g_{\bar{t}\bar{t}}(\bar{r}, \bar{t}) d\bar{t}^2, \\ u^\mu \partial_{x^\mu} &= \sqrt{-g_{\bar{t}\bar{t}}}\, \partial_{\bar{t}} \end{aligned} \tag{13}$$

fixed by†  $\qquad A(\bar{r}, \bar{t}) = A_H(\bar{r}), \qquad g_{\bar{t}\bar{t}}(\bar{r}_b, \bar{t}) = -e^{2\hat{\phi}(\rho_H(\bar{r}_b))} \tag{14}$

for $\bar{r} \in [0, \bar{r}_b]$ with $\bar{r}_b = r_b$ and $(\bar{t} - B_H(\bar{r})) \in [-T, T]$. Our $(r,t)$–coordinates can be transformed into $(\bar{r}, \bar{t})$–coordinates by

$$\bar{t} = t + B_H(r), \quad \bar{r} = r \qquad (\partial_{\bar{t}} = \partial_t, \; \partial_{\bar{r}} = \partial_r - B'_H \partial_t) \tag{15}$$

Moreover, our scalar fields $\rho$‡, $R, U, \Gamma, A$ are the same ones as in [5]–[7] (same notation but different coordinate system). We obtain

$$U = D_t R = e^{-\phi} \partial_t R, \qquad \Gamma = D_{\bar{r}} R = e^{-\lambda/2}(\partial_r - B'_H \partial_t) R \tag{16}$$

from (3),(6). One gains (11) from the analogous formula§ of [5]–[7] by (15).

## 3.  Einstein's equations in (r,t)–coordinates

Given $\rho_H$ and $R_H$ as functions of $r \in [0, r_b]$. Using $\hat{\phi}, \hat{n}, B_H, A_H$ of (4,5) we write Einstein's equations in $(r,t)$–coordinates as follows for $r \neq 0$:

**I)** $(r,t)$-*evolution equations*, which are quasi-linear,

$$\begin{aligned} \partial_t \rho = \alpha^{-1} \cdot (p + \rho) \cdot \Bigg[ & a^2 \cdot B'_H e^{\hat{\phi}(\rho)} \cdot \Gamma \cdot \frac{d\hat{\phi}}{d\rho} \cdot \partial_r \rho - a\, A'_H \cdot \partial_r U \\ & - A'^2_H \cdot \frac{2 U \Gamma}{R} + a A'_H \cdot B'_H e^{\hat{\phi}(\rho)} \cdot \left( \frac{\Gamma^2 - U^2 - 1}{2R} - 4\pi p R \right) \Bigg] \end{aligned} \tag{17}$$

‡ which is called $\epsilon$ in [5]–[7]
§ Similiar results as in [5]–[7] have been obtained by [10].

$$\partial_t R = e^{\hat{\phi}(\rho)} U \tag{18}$$

$$\partial_t U = \alpha^{-1} \cdot \left[ a A'_H \cdot \Gamma^2 \cdot \frac{d\hat{\phi}}{d\rho} \cdot \partial_r \rho - a^2 \cdot B'_H e^{\hat{\phi}(\rho)} \cdot \frac{d\hat{p}}{d\rho} \cdot \Gamma \cdot \partial_r U \right.$$

$$\left. - a A'_H \cdot B'_H e^{\hat{\phi}(\rho)} \cdot \frac{d\hat{p}}{d\rho} \cdot \frac{2U\Gamma^2}{R} + A'^2_H \cdot \Gamma \cdot \left( \frac{\Gamma^2 - U^2 - 1}{2R} - 4\pi p R \right) \right] \tag{19}$$

$$\partial_t \Gamma = \alpha^{-1} \cdot U \cdot \left[ a A'_H \cdot \Gamma \cdot \frac{d\hat{\phi}}{d\rho} \cdot \partial_r \rho - a^2 \cdot B'_H e^{\hat{\phi}(\rho)} \cdot \frac{d\hat{p}}{d\rho} \cdot \partial_r U \right.$$

$$- a A'_H \cdot B'_H e^{\hat{\phi}(\rho)} \cdot \frac{d\hat{p}}{d\rho} \cdot \frac{2U\Gamma}{R}$$

$$\left. + a^2 \cdot B'^2_H e^{2\hat{\phi}(\rho)} \cdot \frac{d\hat{p}}{d\rho} \cdot \left( \frac{\Gamma^2 - U^2 - 1}{2R} - 4\pi p R \right) \right] \tag{20}$$

where $a := 4\pi \hat{n}(\rho) R^2$ and $\alpha := \Gamma \cdot \left[ A'^2_H - a^2 \cdot \left( B'_H e^{\hat{\phi}(\rho)} \right)^2 \cdot \frac{d\hat{p}}{d\rho} \right] e^{-\hat{\phi}(\rho)}$ are technical abbreviations.

**II)** $(r,t)$-*constraints*

$$\Gamma = \frac{4\pi \hat{n}(\rho) R^2}{A'_H} \cdot \left( \partial_r R - B'_H e^{\hat{\phi}(\rho)} U \right), \tag{21}$$

$$\Gamma^2 - U^2 - 1 + \frac{2m}{R} = 0, \tag{22}$$

$$m(r,t) := \int_0^r 4\pi \left[ \rho \partial_r R - B'_H e^{\hat{\phi}(\rho)} \cdot (p+\rho) \cdot U \right](x,t) \cdot R^2(x,t) \, dx; \tag{23}$$

The *scalar field* $m$ is identical with $m$ of Misner (cf [5]–[7]), provided the assumptions of Lemma 3 or Lemma 1 are fulfilled. Misner's $m(\bar{r}, \bar{t})$ denotes the *mass* (total energy) contained in the ball $K(\bar{r}, \bar{t}) := \{(x, \bar{t}) \mid x \in [0, \bar{r}]\}$, where he used the $(\bar{r}, \bar{t})$–coordinates defined in (13),(14).

Einstein's equations in $(\bar{r}, \bar{t})$–coordinates are (A1) together with $\Gamma = \frac{4\pi \hat{n}(\rho) R^2}{A'_H} \cdot \partial_{\bar{r}} R$ and $\Gamma^2 - U^2 - 1 + \frac{2m}{R} = 0$ (see [6]). We transformed these equations via (15) in order to obtain (17–23). Note, that these equations are regular (for $r \neq 0$), since it holds $\alpha \neq 0$ in a neighbourhood of $H$. This follows from (2) with the identities $A'_H = a, B'_H = e^{-\hat{\phi}(\rho)}$ on $H$.

In Lemma 3 we shall see that the constraint equations (21) and (22) will propagate off from $t = 0$.

*Lemma 3.* Given $\rho_H$ as a $C^1$–function and $R_H$ as a $C^2$–function of $r \in [0, r_b]$. Let $U_H, \Gamma_H \in C^1([0, r_b])$ fulfil

$$\Gamma_H = R'_H - U_H \tag{24}$$

$$R_H \Gamma_H^2 - R_H U_H^2 - R_H + 2m_H = 0 \text{ with} \tag{25}$$

$$m_H := \int_0^r 4\pi \left[ \rho_H R'_H - (\hat{p}(\rho_H) + \rho_H) \cdot U_H \right] \cdot R_H^2(x) \, dx \tag{26}$$

on $[0, r_b]$. If $(\rho, R, U, \Gamma)$ is a $C^1$-solution of the evolution equations (17–20) on $]0, r_b] \times [-T, T]$ with

$$\rho = \rho_H, \quad R = R_H, \quad U = U_H, \quad \Gamma = \Gamma_H \text{ on } H, \tag{27}$$

and if $(\rho, R, U, \Gamma)$ can be extended continuously to $r = 0$, then $(\rho, R, U, \Gamma)$ *also fulfils the constraints* (21,22) for all $t \in [-T, T]$, $r \in ]0, r_b]$. Moreover, $R$ is $C^2$ on $]0, r_b] \times [-T, T]$. Thus $(\rho, u^\mu, \mathrm{d}s^2)$ of (6),(7) with (11) fulfil *Einstein's equations*.

The *proof* of this lemma is deferred to Appendix A.

*Remark 4.* Conversely, if Einstein's equations are fulfilled, then (10),(12),(17–20),(24) and (25) are valid for $(\rho_H, R_H, U_H, \Gamma_H) = [(\rho, R, U, \Gamma)]_H$.

## 4. Constructing black hole spacetimes by solving Einstein's equations

At first we shall use Einstein's equations (for perfect fluid) in coordinates, which are – in contrast to our $(r, t)$–coordinates† – *regular at the centre* of symmetry. We shall solve these equations for given data on some auxiliary‡ (spacelike) hypersurface $F$ and thus obtain a perfect fluid spacetime $V_F$ with a *regular centre*. Then we shall construct a further perfect fluid spacetime $V_H$ (as a solution of our $(r, t)$–equations (17–22)) such that $[V_F \cup V_H]$ is a regular extension of $V_F$. The spacetime $V_H$ is determined by data on a hypersurface $H$, which will turn out to be a part of an *event horizon*. We shall finally attach to $[V_F \cup V_H]$ a (unique) vacuum $V_0$ and then gain a *'black hole spacetime'*.

### 4.1. The perfect fluid spacetime $V_F$ being a neighbourhood of the centreline.

*Lemma 5.* Given any smooth 3-manifold $F$ and $C^\infty$-data-functions $\rho_F, U_F$ of $\tilde{r} \in [0, \tilde{r}_0]$ (some $\tilde{r}_0 > 0$). The quantities $\rho_F, U_F$ will become scalar-fields on $F$ (once we have chosen suitable coordinates). We assume for $\rho_F, U_F$

$$\left[\left(\frac{\mathrm{d}}{\mathrm{d}\tilde{r}}\right)^{2n+1} \rho_F\right]_{\tilde{r}=0} = 0, \qquad \left[\left(\frac{\mathrm{d}}{\mathrm{d}\tilde{r}}\right)^{2n} U_F\right]_{\tilde{r}=0} = 0, \text{ for } n = 0, 1, 2, \ldots, \qquad (28)$$

$$\left[\frac{\mathrm{d}U_F}{\mathrm{d}\tilde{r}}\right]_{\tilde{r}=0} < 0 \qquad \text{and} \qquad (29)$$

$$\frac{8\pi}{\tilde{r}} \left(\int_0^{\tilde{r}} \rho_F(x) x^2 \mathrm{d}x\right) < 1 + (U_F(\tilde{r}))^2 \text{ for } \tilde{r} \in [0, \tilde{r}_0]. \qquad (30)$$

Then there exists a unique local (spherically symmetric) $C^\infty$-*solution* $(\mathrm{d}s^2, \rho, u^\mu)$ of Einstein's equations (for perfect fluid) on some 4-manifold $D_F$, where the solution assumes the above (essential) data:

$$\rho = \rho_F, \quad U = U_F \quad on \quad F,$$

where we use $(\tilde{r}, \tilde{t})$-coordinates with $\mathrm{d}s^2$ of type (13) fixed by $F = \{(\tilde{r}, \tilde{t}) \mid \tilde{t} = 0\}$ and $R(\tilde{r}, 0) = \tilde{r}$. Moreover, $F \subset D_F$ is spacelike. Furthermore, there exists a neighbourhood $V_F \subset D_F$ of the *centrepoint* $O$, where the auxiliary condition

$$(D_{\tilde{r}} + D_t)(\Gamma + U) + 4\pi \rho R < 0 \text{ on } V_F \qquad (31)$$

is fulfilled and where $V_F$ is suitably shaped; i.e. $V_F$ is spherical symmetric and the (simply connected) boundary of $V_F$ is spacelike, except for the part between $Q$ and

---

† Our $(r, t)$-equations (17–22) are *singular* at the centre $r = 0$.
‡ We shall *not* give data on the null–cone $H_F$ (see Figure 1), since $H_F$ is a *singular* surface at the centrepoint $0$.

$P$ (see Figure 1); the points $Q$ and $P$ are joined by an outgoing sound–characteristic, whereby $P$ of Figure 1 lies on a null–ray emanating from the centrepoint $O \in F$.

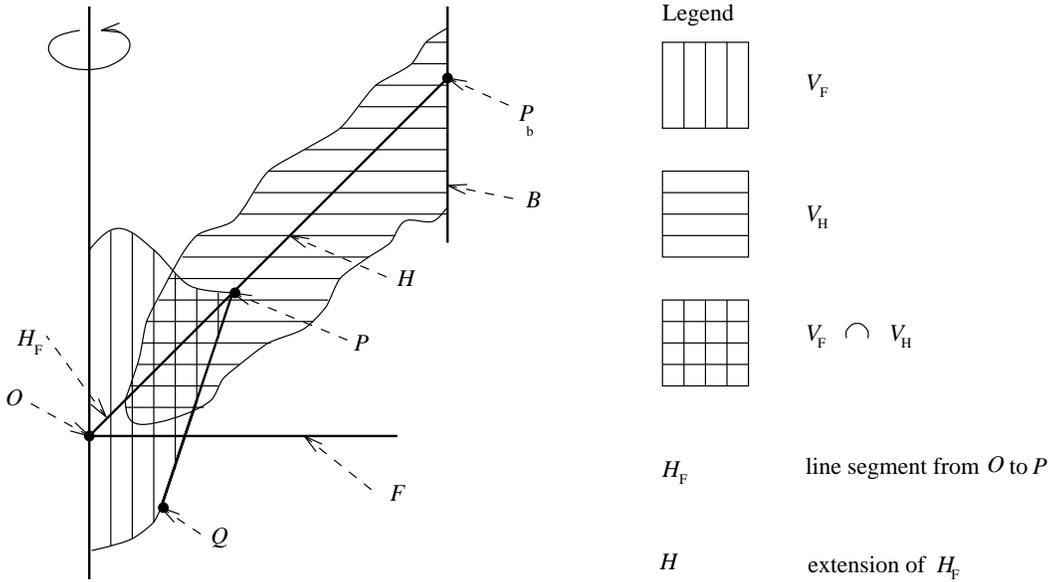

**Figure 1.** The (spherical symmetric) perfect fluid spacetime $V_F \cup V_H$ is determined by data on $F$ and $H$. The spacelike hypersurface $F$ contains the centrepoint $O$. $V_H$ is predicted from $H \setminus$ [neighbourhood of $O$]. The hypersurface $H$ is null (slope 45°).

*Remark 6.* Equation (28) are the necessary and sufficient conditions for $C^\infty$–regularity at the centre $O \in F$. By saying $C^\infty$ ($C^k$, respectively) – without referring to a particular coordinate system – we mean $C^\infty$ ($C^k$, respectively) with respect to *admissible* coordinates. (The polar–coordinates $(\tilde{r}, \tilde{t})$ are non-admissible at $\tilde{r} = 0$.) Furthermore $[R]_F, U_F$ are $C^\infty$ with respect to the *non-admissible* coordinate $\tilde{r}$; but they are *not* $C^1$ (at the centre) in admissible coordinates. However, $[R^2]_F, [R]_F \cdot U_F, \mathrm{d}s^2, \rho_F$ are $C^\infty$ in admissible coordinates. As to further details we refer do Appendix C.

*Proof of Lemma 5.* The data conditions (28–30) are (at $\tilde{t} = 0$) phrased in $(\tilde{r}, \tilde{t})$–coordinates. For Einstein's equations however, we use coordinates, which are – in contrast to the $(\tilde{r}, \tilde{t})$-coordinates (or $(r, t)$-coordinates) – regular at the centre. The equations have a unique (local) spherically symmetric solution (around the centreline) for the Cauchy problem; this has been pointed out (by referring to [11]) in that part of [12], which deals with the domain containing the centreline (and *no* part of the star boundary). In [12] has been used the actual Cauchy–data (the inner metric of $F$ and the extrinsic curvature of $F$). These data can be calculated (in admissible coordinates) from our essential data $\rho_F, U_F$ (see Appendix C). – We obtain (31) from (29) and the following lemma concerning the *regularity*–properties of the *centreline*.

*Lemma 7.* Given a spherically symmetric perfect fluid $C^\infty$–spacetime $V_2$ – such as $V_F$ or $D_F$ (of Lemma 5) – with a regular centreline $C$. We define

$$\Sigma := \{(r, t, \varphi, \vartheta) \mid r \in [0, r_0], t \in [-T, T], \varphi = \varphi_0, \vartheta = \vartheta_0\}$$

for some fixed $\varphi_0, \vartheta_0$ (and sufficiently small $r_0, T$). We consider $\Sigma$ as 2–dimensional submanifold (of $V_2$), which contains the boundary $C$ at $r = 0$. Then it holds

$$\rho \in C^2, \qquad R, \Gamma, U \in C^0(\Sigma), \qquad D_{\tilde{r}}\rho, D_t\rho, \frac{D_{\tilde{r}}U}{\Gamma} \in C^1(\Sigma),$$
$$R = U = 0 \text{ and } \Gamma = 1 \text{ on } C;$$

for $X := \left(\lambda, \partial_r R, \frac{m}{R^2}, D_{\tilde{r}}\Gamma\right)$ and $Y := (D_{\tilde{r}}U, D_tU, D_{\tilde{r}}\Gamma, D_t\Gamma)$ it holds

$$\lim_{r \to 0} [X(r,t)]_{t=0} = (0,1,0,0) \text{ and}$$
$$\lim_{r \to 0} [Y(r,t)]_{t=0} = \lim_{\tilde{r} \to 0} \left[Y(\tilde{r},\tilde{t})\right]_{\tilde{t}=0}, \tag{32}$$

where we expressed $Y$ in the right-hand side of (32) in $(\tilde{r},\tilde{t})$–coordinates used in the context of (30). Furthermore, there exists

$$\lim_{r \to 0} [\partial_r \partial_r R(r,t)]_{t=0}.$$

*Proof.* We use [6] and the equations (A1,A3),(17–23).

*4.2. The perfect fluid spacetime $V_{\mathrm{H}}$ being a neighbourhood of $H \setminus N_0$.*

Let $H_{\mathrm{F}} \subset V_{\mathrm{F}}$ denote the future null–cone of $O \in F$ (see Figure 1).

*Theorem 8.* Given a hypersurface† $H$, being any (sufficently smooth) extension of the null–cone $H_{\mathrm{F}}$ (see Figure 1) and given the *essential data*

$$\rho_{\mathrm{H}}, R_{\mathrm{H}} \text{ on } H \tag{33}$$
$$\rho_{\mathrm{b}} \text{ on the star boundary } B \text{ } (\rho_{\mathrm{b}} \text{ of (2))} \tag{34}$$

where $\rho_{\mathrm{H}}, R_{\mathrm{H}} \in C^3(H \setminus \{0\})$ are extensions of $[(\rho, R)]_{H_{\mathrm{F}}}$ of spacetime $V_{\mathrm{F}}$ (of Lemma 5) and where $(\rho_{\mathrm{H}}, R_{\mathrm{H}})$ fulfil the so-called $(\rho_{\mathrm{H}}, R_{\mathrm{H}})$–conditions (47–53), in particular the 'black hole' condition (50) (see also Remark 9). From the essential data $(\rho_{\mathrm{H}}, R_{\mathrm{H}})$ the non-essential data can be calculated by (43). Let there be given any (sufficently small) neigbourhood $N_0$ of the centrepoint $O$ (on $H_{\mathrm{F}}$) fulfilling $\overline{N_0} \subset V_{\mathrm{F}}$. Then the $(r,t)$-evolution equations (17–20) have a unique local *solution* $(\rho, R, U, \Gamma)$ assuming the above *data*

$$(\rho, R, U, \Gamma) = (\rho_{\mathrm{H}}, R_{\mathrm{H}}, U_{\mathrm{H}}, \Gamma_{\mathrm{H}}) \text{ on } H \setminus N_0 \text{ and } \rho = \rho_{\mathrm{b}} \text{ on } B. \tag{35}$$

The solution $(\rho, R, U, \Gamma)$ defines a *perfect fluid spacetime* $(ds^2, u^\mu, \rho)$ with (6, 7, 11) in some *domain* $V_{\mathrm{H}}$ being a neighbourhood of $H \setminus N_0$. The scalar fields $\rho, U$ and $\Gamma$ are $C^1(V_{\mathrm{H}})$ and $R$ is $C^2(V_{\mathrm{H}})$. Moreover, $H$ ($B$, respectively) is null (timelike, respectively). $H$ will become a piece of an event horizon. Furthermore $[V_{\mathrm{F}} \cup V_{\mathrm{H}}]$ is a perfect fluid spacetime, which is a $C^1$-*extension* of $V_{\mathrm{F}}$.

*Remark 9.* a) The data surface $H$ is a null–surface but *not* a characteristic surface of the system of partial differential equations (17–20), since there do not exist spherically symmetric gravitational waves.

b) The data conditions on $(\rho_{\mathrm{H}}, R_{\mathrm{H}})$ mentioned in Theorem 8 are fulfilled for a large class of data $(\rho_{\mathrm{H}}, R_{\mathrm{H}})$ (see Theorem 14). These conditions consist of several

† see **Remark 9 c)**

*inequalities on H* and several equality relations at the point $P_b$ (of Figure 1), one of these equations at the point $P_b$ is the *black hole* condition (50).

c) The evolution equations used in Theorem 8 are written in our $(r,t)$–coordinates. These coordinates have also to be used for equations (33–35); this means $H = \{(r,t) \mid t = 0, r \in [0, r_b]\}$, $B = \{(r,t) \mid r = r_b\}$ and equation (35) reads

$$\rho(r,0) = \rho_H(r), \ldots, \rho(r_b, t) = \rho_b \quad \text{for} \ r \in [\varepsilon, r_b],$$
$$\text{where } (r,t) = (\varepsilon, 0) \text{ describes the boundary of } N_0 \cap H.$$

*The proof of Theorem 8* is deferred to Appendix B.

*4.3. The construction of the final black hole spacetime.*

*Theorem 10.* a) We give – for details see Lemma 5 – *data* $(\rho_F, U_F)$ on an auxiliary hypersurface $F$; we gain a unique $C^\infty$–*solution* of Einstein's equations (for perfect fluid), where the solution assumes those data $\rho_F, U_F$. The solution is defined in some *domain* $V_F$ (at least); $V_F$ contains a neighbourhood of the *centreline*.

b) We give – for details see Theorem 8 – *data* $(\rho_H, R_H)$ on a hypersurface $H$; we obtain a unique $C^1-$*solution* of the $(r,t)$–evolution equations (17-20) in some *domain* $V_H$ (see Figure 1), whereby the solution assumes those (essential) data $(\rho_H, R_H)$ on $H$ being a null–surface, which will become part of an event horizon $\tilde{H}$. $[V_F \cup V_H]$ is a perfect fluid spacetime, which is a $C^1$–extension of $V_F$ (see Theorem 8). ($R^2$ is $C^2$ on $[V_F \cup V_H]$).

c) We finally *attach* an (up to extension) unique vacuum spacetime $V_0$ to our perfect fluid spacetime $[V_F \cup V_H]$ and thus gain a regular and weakly assymptotically simple $C^1$–spacetime

$$V := [V_F \cup V_H] \cup V_0;$$

$V_0$ is a Schwarzschild spacetime. We *imbed* a smooth *spacelike*, spherically symmetric and simply connected hypersurface $S$ (see Figure 2) in $V$ such that $S \cap [V_F \cup V_H]$ lies in the past of $H$ and that $S \cap V_0$ tends to spacelike infinity of $V_0$. The final spacetime $V$ is then a *'black hole spacetime predicted from S'*, i.e.

(i) the smooth (spherically symmetric) null–surface extension $\tilde{H}$ of $H$ (see Figure 2) is a regular *event horizon*, i.e. $\tilde{H} = \dot{J}^-(\mathcal{J}^+)$ ($\mathcal{J}$ denotes null infinity, notation of [13]); moreover, in the future of $\tilde{H}$ lies a *trapped surface area*, (which lies in parts inside the body,) and a piece of the *Schwarzschild singularity*.;

(ii) The future development of $S$, $D^+(S)$, contains a neighbourhood of the event horizon $\tilde{H}$; the closure (with respect to $V \cup \mathcal{J}$) of $D^+(S)$ contains $\mathcal{J}^+$;

(iii) $D^+(S)$ contains a neighbourhood of the centreline and of the star boundary.

*Proof of Theorem 10.* The vacuum $V_0$ can be attached at the star boundary $B$, since it holds $\rho = \rho_b$ and hence $p = 0$ on $B$. Thus the $V_0$ glued together to the perfect fluid becomes a $C^1$–spacetime (see [12], which refers to the thesis of Kind).

*Theorem 11.* Our spacetime $V$ induces Cauchy data $d_0$ on $S$ (see Theorem 10). From these data it evolves within finite eigentime a unique black hole spacetime $V_{d_0} := V \cap J^+(S)$ including an event horizon and trapped surfaces. The above data $d_0$ on $S$ are regular and *innociuous*, i.e. there are no points on $S$, which belong to the horizon or the trapped surface area (see Figure 2). The formation of the horizon (and the trapped surfaces) is *stable* for spherically symmetric variations of the initial data $d_0$ (on $S$) being in a sufficiently small uniform $C^1$–neighbourhood and for arbitrary variations of the equation of state belonging to a $C^2$–neighbourhood. The above data $d_0$ belong to a class of data, which is at least as large as the class of data posed on $H$ (in Theorem 10).

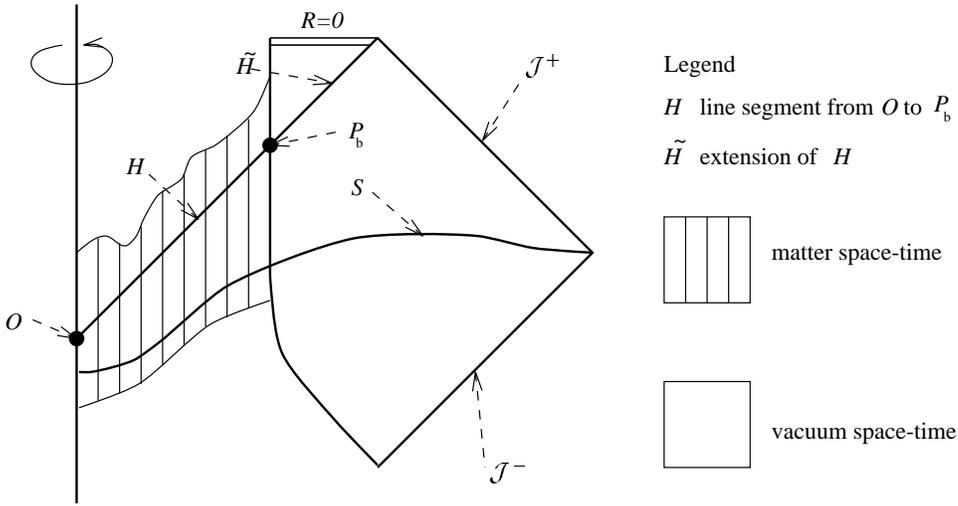

**Figure 2.** The (spherical symmetric) black hole spacetime (with event horizon $\tilde{H}$) is predicted from regular innocuous data an $S$.

*Proof of Theorem 11.* We restrict our attention to the matter part of the spacetime, since the vacuum is uniquely determined by the matter part. Let us give our data $d_0$ on $S$ (and $p = 0$ on $B$). Then we obviuosly obtain a unique solution $[ds^2, u^\mu, \rho]_{d_0}$ on $V_{d_0} := [V_F \cup V_H] \cap J^+(S)$ assuming the above data. This could be proved basically by [12] along with the use of our differentiability classes and coordinates, which are analogous to our $(r, t)$–coordinates. Moreover, this treatment induces *stability* (with respect to spherically symmetric variations): for any data $d$ on $S$, which lies in a sufficiently small uniform $C^1$–neighbourhood of $d_0$, the corresponding $C^1$–solution $[ds^2, u^\mu, \rho]_d$ lies in a given small uniform $C^0$–neighbourhood of $[ds^2, u^\mu, \rho]_{d_0}$. This implies in particular the existence of a null–surface $H_d$ with $2m_d = R_d$ at $\tilde{H}_d \cap B$, since $H_{d_0}$ also has this property ($H_{d_0}$ is our previous $H$, which was the matter part of our horizon). The null–extension of $H_d$ (into the attached vacuum) is an *event horizon*. Moreover, the property of trappedness and collapse is also stable.

## 5. Admissible initial data

Our main theorem (Theorem 10) is a special formulation of the following existence and uniqueness theorem. Given any set of so-called admissible initial data $(\rho_H, R_H, U_H, \Gamma_H)$ then the quasilinear hyperbolic initial–boundary problem (17–20) with (35) yields a spacetime $V_H$ with the properties stated in Theorem 8 (, provided the data are extensions of $[(\rho, R, U, \Gamma)]_{H_F}$). Here we call the functions $(\rho_H, R_H, U_H, \Gamma_H)$ of $r \in [0, r_b]$

*admissible initial data*, if they fulfil

| | | |
|---|---|---|
| constraint equations: | $R'_H (R'_H - 2U_H) = 1 - 2m_H/R_H, \quad \Gamma_H = R'_H - U_H$ | (36) |
| | (equivalent to (24,25)); | |
| non-negative pressure: | $\hat{p}(\rho_H) > 0$ with equality relation in $r_b$; | (37) |
| collapse: | $U_H < 0$ on $]0, r_b]$; | (38) |
| light escape: | $R'_H > 0$ on $[0, r_b[$; | (39) |
| black hole condition: | $2m_H(r_b) = R_H(r_b)$; | (40) |
| corner condition: | right-hand side of (17) equals 0 at $r = r_b$ | (41) |
| | (for $(\rho, R, U, \Gamma) = (\rho_H, R_H, U_H, \Gamma_H)$) ; | |
| regular centre: | $R_H(0) = 0, R'_H(0) = 1, U_H(0) = 0$. | (42) |

In general, one can prescribe any two of the four functions $(\rho_H, R_H, U_H, \Gamma_H)$ and obtain the others from the constraint equations. In practice, the most simple† way seems to be to prescribe $\rho_H$ and $R_H$, which we therefore called *essential data* in the previous sections. The non-essential data $U_H$ and $\Gamma_H$ can be calculated from the *constraints* as

$$U_H = \mathcal{U}(\rho_H, R_H) \text{ and } \Gamma_H = R'_H - U_H, \tag{43}$$

where the operator $\mathcal{U}$ is defined by

$$\mathcal{U}(\rho_H, R_H) : [0, \infty[ \to \mathbb{R}, \ r \mapsto \frac{e^{-\chi(r)}}{R_H(r)} \cdot \int_0^r \left[ \frac{(R'_H)^2 - 1}{2} + q_H \cdot R_H \right] \cdot e^{\chi(x)} dx \tag{44}$$

$$\text{with } q_H := R''_H + 4\pi \rho_H R_H \text{ and} \tag{45}$$

$$\chi(r) := \int_0^r \left[ \frac{q_H + 4\pi \hat{p}(\rho_H) R_H}{R'_H} \right] dx. \tag{46}$$

Since we shall prescribe the essential data $(\rho_H, R_H)$ and calculate the other data by (43), we shall now rewrite the conditions (36–42) as conditions on $\rho_H$ and $R_H$.

*Lemma 12.* Let $\rho_H \in C^1$ and $R_H \in C^2$ for $r \in [0, r_b]$. Then $(\rho_H, R_H, U_H, \Gamma_H)$ with $U_H \in C^1([0, r_b])$ are *admissible initial data* (i.e. fulfil (36–42)) if and only if $U_H = \mathcal{U}(\rho_H, R_H), \Gamma_H = R'_H - U_H$ and the following $(\rho_H, R_H) - conditions$ are fulfilled:

| | | |
|---|---|---|
| $\rho > \rho_b$ on $[0, r_b[ \ \rho(r_b) = \rho_b$ | (non-negative pressure), | (47) |
| $\mathcal{U}(\rho_H, R_H) < 0$ in $]0, r_b]$ | (collapse), | (48) |
| $R'_H > 0$ on $[0, r_b[$ | (light escape), | (49) |
| $R'_H(r_b) = 0$ | (black hole condition), | (50) |
| $q'_H(r_b) = 8\pi \cdot \rho_b \cdot \mathcal{U}(\rho_H, R_H)(r_b)$ | (corner condition), | (51) |
| $q_H(r_b) = 0$ | (see Remark 13), | (52) |
| $R_H(0) = 0, R'_H(0) = 1$ | (regular centre of symmetry). | (53) |

Before we proof this lemma, let us exhibit a few properties of these $(\rho_H, R_H)$–conditions.

---

† For given $(\rho_H, R_H)$ the quantity $[U_H \cdot R_H]$ is a solution of a *linear* differential equation being equivalent to the first constraint equation in (36).

*Remark 13.* a) Note, that (48,49) are inequalities on $H$, whereas (50,51) are equality relations at the point $H \cap B$. The corner condition (51), which is equivalent to (41) is a technical condition ensuring a convenient degree of differentiability.

b) The inequalities (47–49) are *stable* under small variations (of $(\rho_H, R_H)$) keeping the black hole condition (50) and $\rho(r_b) = \rho_b$ fixed. This follows from the continuity of the operator $\mathcal{U}$.

c) The black hole condition (50) is equivalent with $\left[\frac{2m}{R}\right]_{H \cap B} = 1$, provided it holds (36, 48,49).

d) Equation (49) *means* $(D_{\tilde{r}} + D_t)R > 0$ *on* $H \setminus B$, where $H$ is – among other things – a future–lightray. In general, it holds for *any* connected future–lightray $L \subset [V_F \cup V_H]$ the following: $L$ *escapes*‡ *to* $\mathcal{J}^+$, *if and only if* $(D_{\tilde{r}} + D_t)R > 0$ *on* $L \setminus B$, provided L hits the star boundary $B$ and $\rho > 0$ on $V_F \cup V_H \cup B$, (which holds for sufficiently small $V_F$ and $V_H$.) The escape–condition can be proven by [6] and the fact that $(D_{\tilde{r}} + D_t)R$ is strictly positive (negative, respectively) on $B \cap V_0 \cap [\mathcal{J}^-(H) \setminus H]$ $\quad$ $(B \cap V_0 \cap [\mathcal{J}^+(H) \setminus H]$, respectively).

e) Equation (49) is equivalent with $\left[\frac{2m}{R}\right]_{H \setminus B} < 1$, provided it holds (36,48,53). Furthermore, (49) prevents $\Gamma = 0$, which would make the evolution equations (17–20) singular. Moreover, $\Gamma < 0$ would mean a 'neck' (see [5]).

f) Equation (52) implies that the integrand of $\chi$ of (46) is – despite (50) – continuous. Thus $\mathcal{U}_H(\rho_H, R_H)$ is $C^1$ also at $r = r_b$ (cf (46)). Conversely, the first constraint equation of (36) and $R'_H(r_b) = 0$, which is a consequence of (36–42), imply (52).

*Proof of Lemma 12.* At first we show the equivalence of the constraint equations (36) with (43) by rewriting the first equation of (36) as

$$R'_H R_H U_H = m_H + R_H \cdot \left((R'_H)^2 - 1\right)/2 \tag{54}$$

with $m'_H = 4\pi \left(\rho_H R'_H - (\hat{p}(\rho_H) + \rho_H) U_H\right) R_H^2$ of (26). Then $\left[\frac{d}{dr}(54)\right]/R'_H$ becomes a *linear* differential equation in the new unknown $W := [R_H \cdot U_H]$. For initial value $W(0) = 0$ it has a unique solution $W$, which is explicitly writable as an integral. Indeed, we defined in (44) our $\mathcal{U}(\rho_H, R_H)$ as $W/R_H$; this yields the

$$\text{equivalence of (36) with (43) on } ]0, r_b[ \text{ at least.} \tag{55}$$

If (36–42) are fulfilled with $U_H \in C^1([0, r_b])$ then $U_H(0) = 0 = \mathcal{U}(\rho_H, R_H)(0)$ by Hospital's rule. Because of $U'_H = (\mathcal{U}(\rho_H, R_H))'$ on $]0, r_b[$ there exists

$$\lim_{r \to r_b} (\mathcal{U}(\rho_H, R_H))'(r) = \lim_{r \to r_b} \left[-\left(\chi' + \frac{R'_H}{R_H}\right)\mathcal{U}(\rho_H, R_H) + \frac{(R'_H)^2 - 1}{2R_H} + q_H\right](r).$$

Together with $\lim_{r \to r_b} (\mathcal{U}(\rho_H, R_H))(r) = U_H(r_b) \in ]0, \infty[$ it follows the existence of (a finite) $\lim_{r \to r_b} \chi'(r) = \chi'(r_b)$. This implies $q_H(r_b) = 0$ due to (50), which follows from (36), (40) and $(R'_H - 2U_H)(r_b) \neq 0$.

Conversely, if it holds (43) and (47–53) then it follows

$$U_H(0) = 0, (36) \text{ for } r = 0 \text{ and } U_H \in C^1([0, r_b])$$

from (44), where we use Hospital's rule at $r = 0$ ($r = r_b$, respectively) bearing in mind (53) ((52), respectively). From this and (55) we gain (36) on $[0, r_b[$, which can be extended to $[0, r_{\mathrm{b}}]$ by the continuity of the quantities in (36). Then the black hole condition (40) follows from (50).

---

‡ If $L$ hits $B \cap H$, then $H$ is the unique extension of $L$, i.e. $H$ 'escapes to $\mathcal{J}^+$' in the sense that there exists an affine parameter, which tends to *infinity* (when moving towards the future).

It remains to show the equivalence of the different formulations of the corner condition. The evaluation of (41) just yields (51), where we used (17) for $(\rho, R, U, \Gamma) = (\rho_{\mathrm{H}}, R_{\mathrm{H}}, U_{\mathrm{H}}, \Gamma_{\mathrm{H}})$, $a = A'_{\mathrm{H}}, B'_{\mathrm{H}} \mathrm{e}^{\hat{\phi}(\rho)} = 1, p = 0$, (22), (40), and (43, 46)' with (50) and the rule of l'Hospital for $q_{\mathrm{H}}/R'_{\mathrm{H}}$. This finishes the proof.

We shall now show that the $(\rho_{\mathrm{H}}, R_{\mathrm{H}})$-conditions are not too restrictive. At first, we choose $\rho_{\mathrm{H}}$ as *free initial data* (with (47)). Then the amount of freedom left to the choice of $R_{\mathrm{H}}$ amounts to the cardinality of $C^1\left([0,1], ]0, -\infty[\right)$ (at least).

*Theorem 14.* Let $\rho_{\mathrm{H}}$ be given with (47) and

$$16\pi \left(\max_{[0, r_{\mathrm{b}}]} \{\rho_{\mathrm{H}}(r)\}\right) r_{\mathrm{b}}^2 < 1.$$

Then for any auxiliary $C^1$-function $h : [0,1] \to ]0, -\infty[$ we can construct an 'admissible' $R_{\mathrm{H}} = R_{\mathrm{H}}^h$ such that $(\rho_{\mathrm{H}}, R_{\mathrm{H}})$ fulfil the $(\rho_{\mathrm{H}}, R_{\mathrm{H}})$-conditions (47–53), i.e. $(\rho_{\mathrm{H}}, R_{\mathrm{H}}, U_{\mathrm{H}}, \Gamma_{\mathrm{H}})$ are admissible initial data, where $(U_{\mathrm{H}}, \Gamma_{\mathrm{H}})$ are defined by (43). Furthermore, our construction yields $R_{\mathrm{H}}^h \neq R_{\mathrm{H}}^{\tilde{h}}$ for $h \neq \tilde{h}$.

We shall describe the main ideas of our *proof*. Since we are concerned with quantities (such as $\rho_{\mathrm{H}}, R_{\mathrm{H}}, U_{\mathrm{H}}, \ldots$) defined on $H$ *only*, we shall drop the index $H$ and bear in mind that these quantities depend on $r$ only. From (44) one sees that

$$q = R'' + 4\pi\rho R < 0, \tag{56}$$

$R' < 1$, (53) and (49) are sufficient for $U < 0$.

Therefore we do not choose $R$ with (48), (49) etc., but we choose a negative $q \in C^0$, which defines $R := \mathcal{R}(q)$ as the solution of

$$R'' + 4\pi\rho R = q \qquad \text{with } R(0) = 0 \text{ and } R'(0) = 1. \tag{57}$$

The operater $\mathcal{R} : C^0 \to C^2$ is continuous if one endowes the range of $\mathcal{R}$ by the $C^1$-norm (uniform topology).

In order to gain an $R := \mathcal{R}(q)$ as in Theorem 8 we choose $q$ to be $q = q_0$ on a sufficiently small part of $H_{\mathrm{F}}$, i.e. for $r \in [0, \varepsilon]$; $q_0$ is defined as

$$q_0 := [\text{left-hand side of (31)}]_{H_{\mathrm{F}}} < 0. \tag{58}$$

Moreover, we postulate $q$ to be $C^1$ near $r_{\mathrm{b}}$, since $q'(r_{\mathrm{b}})$ occurs in the corner condition (51).

*Lemma 15.* Let $\rho$ and $h$ be given as in Theorem 14. Then a real parameter $\sigma$ and a positive parameter $\tau$ exist such that

$$q := q_{(\sigma, \tau)} := \kappa_\tau (q_h + \eta_\sigma) \tag{59}$$

induces (in the sense of Theorem 14) an admissible $R := \mathcal{R}(q)$ ($\mathcal{R}$ is defined around (57)). Here we define $q_h$ as a continuous function with $q_h(r) := h\left(\frac{r - 3\varepsilon}{r_{\mathrm{b}} - 6\varepsilon}\right)$ for $r \in [3\varepsilon, r_{\mathrm{b}} - 3\varepsilon]$, $q_h = q_0$ on $[0, \varepsilon]$, $q_h < 0$ on $[0, r_{\mathrm{b}} - \varepsilon[$ and $q_h \equiv 0$ on $[r_{\mathrm{b}} - \varepsilon, \infty[$, where $q_0$ was defined in (58). The auxiliary functions $\kappa_\tau$ and $\eta_\sigma$ are defined below.

*Remark 16.* The function $q_h$ of Lemma 15, *which determines* $R := \mathcal{R}(q_h)$, carries the information of $h$. Moreover, if the assumptions of Theorem 8 are fulfilled, then $R := \mathcal{R}(q_h)$ becomes an extension of $[R]_{H_{\mathrm{F}}}$ by choosing $q_0 := [\text{left-hand side of } (31)]_{H_{\mathrm{F}}} < 0$, $q_0 \in C^0([0, \varepsilon])$. Furthermore, $q_h$ is $C^1$ near $r_\mathrm{b}$.

*Scetch of the proof of Lemma 15 including the definitions of $\kappa_\tau$ and $\eta_\sigma$.*

$$\text{The parameter } \tau \text{ (cf (59)) is used for positioning the (first)} \tag{60}$$
$$\text{zero of } R' = \left[\mathcal{R}\left(q_{(\sigma,\tau)}\right)\right]' \text{ at } r = r_\mathrm{b}.$$

When we increase $\tau$ (decrease, respectively) then the zero of $R'$ decreases (increases, respectively). $\kappa_\tau$ is defined as $\kappa_\tau = 1 + (\tau - 1)\kappa$, where $\kappa : [0, \infty[ \to [0, 1]$ is any fixed $C^\infty$–function with support $\subset [\varepsilon, r_\mathrm{b} - \varepsilon]$ and $\kappa \equiv 1$ on $[2\varepsilon, r_\mathrm{b} - 2\varepsilon]$. We define a zero–searcher $\zeta$ such that $\zeta(q)$ is the first (positive) zero of the derivative of $\mathcal{R}(q)$, provided $q$ lies in a (sufficiently small) neighbourhood of a negative function†.

It can be shown that $\tau_1, \tau_2$ exist with

$$\zeta\left(\kappa_{\tau_2} \tilde{q}\right) < r_\mathrm{b} < \zeta\left(\kappa_{\tau_1} \tilde{q}\right) \text{ and } \tau_1 < \tau_2 \tag{61}$$

for all $\tilde{q}$ of a $C^0$–neighbourhood of $q_h$. Moreover, the operator $\zeta$ is continuous there. Let such a $\tilde{q}$ be given. Then we can find (according to the intermediate value theorem) a $\kappa_\tau$ (with $\tau \in [\tau_1, \tau_2]$), which does its job: $\zeta(\kappa_\tau \cdot \tilde{q}) = r_\mathrm{b}$.

The other family of auxiliary functions $\eta_\sigma$ *controls the derivative of* $q_{(\sigma,\tau)}$ at $r_\mathrm{b}$ with the aim to fulfil the corner condition (51). Unfortunately, a variation of $q'(r_\mathrm{b}) = \eta'_\sigma(r_\mathrm{b}) = \sigma$ violates the requirement $q < 0$ and also handicaps the aim concerning the choice of $\tau$ (see (60)). The first item does not matter provided $|\eta_\sigma|$ is small enough, say $|\eta_\sigma| < \delta_q$ for some constant $\delta_q < \varepsilon$.

We define $\eta_\sigma : [0, \infty[ \to [-\delta_q, \delta_q]$ by

$$\eta_\sigma(r) := \delta_\sigma \, \mathrm{sign}(\sigma) \cdot \tilde{\eta}\left(\frac{r - r_\mathrm{b}}{\delta_\sigma} |\sigma|\right)$$

with $\delta_\sigma := \min\{\delta_q, \varepsilon|\sigma|\}$, where $\tilde{\eta} : [0, \infty[ \to [-1, 1]$ is any $C^1$–function with $\tilde{\eta}'(0) = 1$ and support $\subset [-1, 1]$.

It can be shown that $\{\tilde{q} \in C^0 \mid \tilde{q} = q_h + \eta_\sigma \text{ with } \sigma \in \mathbb{R}\}$ lies in a neighbourhood of $q_h$ such that there exist $\tau_1, \tau_2$ with (61). We define the set of *'black hole'- parameters*

$$E := \left\{(\sigma, \tau) \in \mathbb{R} \times [\tau_1, \tau_2] \mid R'(r_\mathrm{b}) = 0 \text{ for } R := \mathcal{R}\left(q_{(\sigma,\tau)}\right) \text{ (cf (59))}\right\}$$

(cf condition (50)). For any given $\sigma \in \mathbb{R}$ the set $E$ contains one $(\sigma, \tau)$ at least; thus $(\sigma, \tau) \in E$ means *no restriction on* $\sigma$. Up to now, we know that $(\rho, R)$ with $R := \mathcal{R}\left(q_{(\sigma,\tau)}\right)$ fulfil the $(\rho_\mathrm{H}, R_\mathrm{H})$-conditions, if $(\sigma, \tau) \in E$, provided the corner condition (51) is satisfied for $q_\mathrm{H} = q_{(\sigma,\tau)}$ and $U_\mathrm{H} = \mathcal{U}(\rho, R)$. Due to $q'_{(\sigma,\tau)}(r_\mathrm{b}) = \sigma$ the corner condition is fulfilled, if it holds

$$\sigma = 8\pi \rho_\mathrm{b} U_\mathrm{b}(\sigma, \tau), \tag{62}$$

where the function $U_\mathrm{b} : E \to \mathbb{R}$ is defined as $U_\mathrm{b}(\sigma, \tau) := \left[\mathcal{U}\left(\rho, \mathcal{R}\left(q_{(\sigma,\tau)}\right)\right)\right](r_\mathrm{b})$. Thus it remains to show the existence of such a $(\sigma, \tau) \in E$, which fulfills (62). Since $U_\mathrm{b}$ is uniformly bounded and continuous‡, we have to look for these $(\sigma, \tau) \in E$ with (62) only in some compactum $[\sigma_1, \sigma_2] \times [\tau_1, \tau_2]$. We show that such a $(\sigma, \tau)$ exists by the intermediate value theorem along with some technical arguments. Thus we have found an admissible $R = \mathcal{R}\left(q_{(\sigma,\tau)}\right)$.

† Then $(\mathcal{R}(q))'$ is strictly monotonically decreasing, which makes $\zeta$ well-defined and continuous.

It is not hard to see that $q_h \neq q_{\tilde{h}}$ lead to admissible $R^h \neq R^{\tilde{h}}$.

**Acknowledgments**

We thank Professor B G Schmidt for drawing our attention to this topic and for useful discussions. We are grateful to Professor H–J Seifert for several helpful remarks. We also like to thank Dr A D Rendall and Dr H Friedrich for useful discussions.

**Appendix A.**

*Proof of Lemma 3.* For $r \neq 0$ it holds that a $C^1$-solution $(\rho, R, U, \Gamma)$ of (17–20) is also a $C^1$-solution of

$$\begin{aligned}
\partial_{\bar{t}} \rho &= -e^{\hat{\phi}(\rho)} (p + \rho) \left( \frac{4\pi \cdot \hat{n}(\rho) \cdot R^2}{A'_{\mathrm{H}} \cdot \Gamma} \cdot \partial_{\bar{r}} U + \frac{2U}{R} \right) \\
\partial_{\bar{t}} R &= e^{\hat{\phi}(\rho)} U \\
\partial_{\bar{t}} U &= e^{\hat{\phi}(\rho)} \left( \frac{4\pi \cdot \hat{n}(\rho) \cdot R^2}{A'_{\mathrm{H}}} \cdot \Gamma \cdot \frac{\mathrm{d}\hat{\phi}}{\mathrm{d}\rho} \cdot \partial_{\bar{r}} \rho + \frac{\Gamma^2 - U^2 - 1}{2R} - 4\pi p R \right) \\
\partial_{\bar{t}} \Gamma &= e^{\hat{\phi}(\rho)} \cdot \frac{4\pi \cdot \hat{n}(\rho) \cdot R^2}{A'_{\mathrm{H}}} \cdot U \cdot \frac{\mathrm{d}\hat{\phi}}{\mathrm{d}\rho} \cdot \partial_{\bar{r}} \rho \ .
\end{aligned} \quad (A1)$$

(to be proved via (15) assuming (2)).

**Remark.** Note that (A1) turns out to be Misner's evolution equations (with $\frac{\mathrm{d}A}{\mathrm{d}r}$ chosen as $A'_{\mathrm{H}}$), provided one formally substitutes his $-2m/R$ by $\Gamma^2 - U^2 - 1$.

We now shall need $\partial_{\bar{r}} \partial_{\bar{t}} R = \partial_{\bar{t}} \partial_{\bar{r}} R$ and hence prove $R \in C^2$ by defining a scalar field $\tilde{R}$ (unique up to an integration constant) with

$$\partial_{\bar{r}} \tilde{R} = \frac{A'_{\mathrm{H}}}{4\pi \cdot \hat{n}(\rho) \cdot R^2} \cdot \Gamma \ \text{and} \ \partial_{\bar{t}} \tilde{R} = e^{\hat{\phi}(\rho)} U \ , \quad (A2)$$

where $\partial_{\bar{t}} \left( \frac{A'_{\mathrm{H}}}{4\pi \cdot \hat{n}(\rho) \cdot R^2} \cdot \Gamma \right) = \partial_{\bar{r}} \left( e^{\hat{\phi}(\rho)} U \right)$ because of (A1). Thus (A2) implies $\tilde{R} \in C^2$. From (15) we get for $r \neq 0$

$$\partial_r \tilde{R} = \frac{A'_{\mathrm{H}}}{4\pi \cdot \hat{n}(\rho) \cdot R^2} \cdot \Gamma + B'_{\mathrm{H}} e^{\hat{\phi}(\rho)} U \ , \quad (A3)$$

thus $\left[ \partial_r \tilde{R} \right]_{\mathrm{H}} = R'_{\mathrm{H}}$ for $r \neq 0$ because of (4,5,24). This and fixing $\tilde{R}$ by $\tilde{R}(\bar{r}_0, B_{\mathrm{H}}(\bar{r}_0)) = R_{\mathrm{H}}(\bar{r}_0)$ (for any fixed $\bar{r}_0 \in \,]0, r_{\mathrm{b}}]$) gives $R = \tilde{R}$ on $H$ (for $r \neq 0$), which (together with $\partial_t R = \partial_t \tilde{R}$) yields $R = \tilde{R} \in C^2$ in $]0, r_b] \times [-T, T]$. From (A3) and $R = \tilde{R}$ it follows (21).

---

‡ In order to proof this we define the continuous operator $\mathcal{T} : \left\{ f \in C^1 \text{ near } r_{\mathrm{b}} \mid f(r_{\mathrm{b}}) = 0 \right\} \to C^0$ as $(\mathcal{T}(f))(r) := \frac{f(r)}{r - r_{\mathrm{b}}}$. Then the continuity of $E \to C^0, (\sigma, \tau) \mapsto \chi' = \frac{\mathcal{T}\left(q_{(\sigma,\tau)}\right) + 4\pi \cdot \mathcal{T}(p) \cdot \mathcal{R}\left(q_{(\sigma,\tau)}\right)}{\mathcal{T}\left(\left(\mathcal{R}(q_{(\sigma,\tau)})\right)'\right)}$ can be shown.

Analogously to the construction of $\tilde{R}$, we define $\tilde{m} \in C^2$ (unique up to a constant) by

$$\partial_{\bar{r}} \tilde{m} = v_{\bar{r}} := 4\pi \rho R^2 \cdot \frac{A'_H}{4\pi \cdot \hat{n}(\rho) \cdot R^2} \cdot \Gamma \quad \text{and} \quad \partial_{\bar{t}} \tilde{m} = v_{\bar{t}} := -4\pi p R^2 U e^{\hat{\phi}(\rho)} \tag{A4}$$

for $\bar{r} \in ]0, r_b], (\bar{t} - B_H(\bar{r})) \in [-T, T]$, where $\partial_{\bar{r}} v_{\bar{t}} = \partial_{\bar{t}} v_{\bar{r}}$, which follows from $\partial_{\bar{r}} \partial_{\bar{t}} R = \partial_{\bar{t}} \partial_{\bar{r}} R$ and (A1). Now we consider the scalar field $\tilde{m}$ in $(r,t)$-coordinates; we use (15),(A4),(21) and obtain for $r \neq 0$:

$$\partial_r \tilde{m} = \partial_r m, \qquad (m \text{ of } (23)) \tag{A5}$$

$$\tilde{m}(r, 0) = \tilde{m}(r_0, 0) + \int_{r_0}^{r} \partial_r m(x, 0) \, dx, \tag{A6}$$

$$\tilde{m}(r, t) = \tilde{m}(r, 0) + \int_{0}^{t} [v_{\bar{t}}(\bar{r}, \bar{t})]_{(\bar{r}, \bar{t}) = (r, x + B_H(r))} \, dx. \tag{A7}$$

Furthermore, $\partial_r m(r, 0) = m'_H$ is continuous on $[0, r_b]$ (express again $\partial_r R$ by $\Gamma$ of (21)); this and (A6) implies that $\lim_{r \to 0} \tilde{m}(r, 0)$ exists; now, $\tilde{m}$ was fixed up to a constant; we choose this constant such that $\lim_{r \to 0} \tilde{m}(r, 0) = 0$. This and the fact that $v_{\bar{t}}$ tends continuously towards 0, implies $\lim_{r \to 0} \tilde{m}(r, t) = 0$ (using (A7)). From this and from $\lim_{r \to 0} m(r, t) = 0$ ($m$ of (23)) we gain the identity $\tilde{m} = m$ by (A5).

Equations (A1) and (A4) along with $\tilde{m} = m$ implies $\partial_{\bar{t}} \left[ \Gamma^2 - U^2 - 1 + \frac{2m}{R} \right] = -e^{\hat{\phi}(\rho)} \frac{U}{R} \cdot \left[ \Gamma^2 - U^2 - 1 + \frac{2m}{R} \right]$. This and $\left[ \Gamma^2 - U^2 - 1 + \frac{2m}{R} \right]_H = 0$ (due to (25)) implies that the *constraint* (22) *is valid*.

Due to (A4) Misner's mass (cf [5]–[7]) can be identified as $\tilde{m}$, i.e. our $m$ and Misner's $m$ are the same scalar fields. This, (A1) and the constraints (21,22) imply Misner's equations (with $\frac{dA}{dr}$ chosen as $A'_H$), as stated in the remark under (A1). Thus Einstein's equations hold. – Note, that our evolution system (A1) appears in a form, to which Wesson and de Ponce do not object in [14].

**Appendix B.**

*Proof of Theorem 8.* Courant and Hilbert [15, p 425, p 476, pp 461 ff] proved the existence and uniqueness of the solution of the following mixed problem for a quasilinear system (in two variables). Let $\boldsymbol{u} = (R, \rho, U, \Gamma)$ be the vector of the unknowns. Then one can write (17–20) as

$$\partial_t \boldsymbol{u} + \mathbf{M} \cdot \partial_r \boldsymbol{u} = \boldsymbol{f}$$

where the $4 \times 4$-Matrix $\mathbf{M}$ and $\boldsymbol{f}$ depend on $(r, t, \boldsymbol{u}(r, t))$. The eigenvalues of $\mathbf{M}$ are $\lambda_{1,2} = 0$ and

$$\lambda_{3,4} = \frac{\pm \sqrt{\frac{d\hat{p}}{d\rho}} 4\pi n R^2 e^{\hat{\phi}(\rho)}}{A'_H \mp \sqrt{\frac{d\hat{p}}{d\rho}} 4\pi n R^2 B'_H e^{\hat{\phi}(\rho)}}.$$

We can choose as associated left eigenvectors

$$\boldsymbol{v}_1 = (0, 1, 0, 0), \boldsymbol{v}_2 = (0, 0, -U, \Gamma) \text{ and } \boldsymbol{v}_{3,4} = \left( \sqrt{\frac{d\hat{p}}{d\rho}}, 0, \pm \frac{p + \rho}{\Gamma}, 0 \right),$$

i.e. $\lambda_i \boldsymbol{v}_i = \boldsymbol{v}_i \cdot \mathbf{M}$ for $i = 1, 2, 3, 4$. Our system is *hyperbolic*, since these eigenvectors are linear independent for $\boldsymbol{u} = \boldsymbol{u}_\mathrm{H}$. The canonical functions $g_1 := \boldsymbol{v}_1 \cdot \boldsymbol{u} = R$ and $g_2 := \boldsymbol{v}_2 \cdot \boldsymbol{u} = \Gamma^2 - U^2$ propagate along the stream lines, while

$$g_{3,4} := \boldsymbol{v}_{3,4} \cdot \boldsymbol{u} = \sqrt{\frac{\mathrm{d}\hat{p}(\rho)}{\mathrm{d}\rho}} \cdot \rho \pm \frac{p + \rho}{\Gamma} \cdot U$$

propagate along out-/in-going sound characteristics (cf [16]). So our mixed problem is well-posed [15, p 472] if we give initial data $\boldsymbol{u}(r, 0) = \boldsymbol{u}_\mathrm{H}(r)$ and boundary values for $g_4$, namely

$$g_4(r_\mathrm{b}, t) = -g_3(r_\mathrm{b}, t) + 2\sqrt{\frac{\mathrm{d}\hat{p}(\rho_\mathrm{b})}{\mathrm{d}\rho}} \cdot \rho_\mathrm{b}.$$

These data mean $\rho(r_\mathrm{b}, t) \equiv \rho_\mathrm{b}$ (for $t \in [-T, T]$), i.e. vanishing pressure on the boundary of the star. This initial-boundary value problem has a solution with Lipschitz continuous first derivatives, provided $\mathbf{M}, \boldsymbol{f}, \lambda_i, \boldsymbol{v}_i$ (and $\boldsymbol{u}_\mathrm{H}$ respectively) have Lipschitz continuous first derivatives with respect to $t, r, \boldsymbol{u}$ (and $r$ respectively).

We have to show the *uniqueness* of the solution in $V_\mathrm{F} \cap V_\mathrm{H}$, i.e. that part of the spacetime in figure 1, which is marked twice (by parallel lines in horizontal as well as in vertical direction). The boundary of $V_\mathrm{F} \cap V_\mathrm{H}$ is such that the slopes of its tangents are not steeper than those of the sound characteristic, which is the fastest characteristic. This yields

$$[V_\mathrm{F} \cap V_\mathrm{H}] \subset D_\mathrm{s}(H_1), \qquad H_1 := H \cap [V_\mathrm{F} \cap V_\mathrm{H}] \tag{B1}$$

where $D_\mathrm{s}$ denotes the (future and past) domain of dependence with respect to the sound characteristic. On the other hand, we have solutions of the evolution equations (e.g. (A1) for $r > 0$) together with the constraint equations (21,22), since Lemma 3 is extendable to any simple connected domain with simple connented stream lines by extending $m$ as $\tilde{m}$. These solutions of Einstein's equations exist in $V_\mathrm{F}$ as well as in $V_\mathrm{H}$ and coincide on $H \cap V_\mathrm{F} \cap V_\mathrm{H}$. This and (B1) yields coincidence of the solution in $V_\mathrm{F} \cap V_\mathrm{H}$. This finishes the proof of Theorem 8.

## Appendix C.

*Scetch of the proof of Remark 6.* It has been pointed out in [12] that (28) is a necessary condition for $C^\infty$-regularity at the centre. We show that these conditions are sufficient: the actual data of our Cauchy-problem in Lemma 5 are the inner metric $g^\mathrm{F}_{\alpha\beta}$ of $F$ ($\alpha, \beta$ runs $1, 2, 3$) and the extrinsic curvature $K^\mathrm{F}_{\alpha\beta}$ of $F$. They can be calculated from our (essential) data $\rho_\mathrm{F}, U_\mathrm{F}$ as

$$g^\mathrm{F}_{\alpha\beta} = \frac{E - 1}{\tilde{r}^2} x_\alpha x_\beta + \delta_{\alpha\beta}, \quad K^\mathrm{F}_{\alpha\beta} = \left(\frac{EU'_\mathrm{F}}{\tilde{r}^2} - \frac{U_\mathrm{F}}{\tilde{r}^3}\right) x_\alpha x_\beta - \frac{U_\mathrm{F}}{\tilde{r}} \delta_{\alpha\beta} \tag{C1}$$

with

$$E := g_{\tilde{r}\tilde{r}}(\tilde{r}, 0) = \left(U_\mathrm{F} \cdot U_\mathrm{F} + \frac{8\pi}{\tilde{r}} \int_0^{\tilde{r}} \rho_\mathrm{F}(x) x^2 \, \mathrm{d}x + 1\right)^{-1}, \tag{C2}$$

where we used (13), $R(\tilde{r}, 0) = \tilde{r}$ and the coordinates†

$$x_1 = \tilde{r} \cos\varphi \sin\vartheta, \qquad x_2 = \tilde{r} \sin\varphi \sin\vartheta, \qquad x_3 = \tilde{r} \cos\vartheta$$

† These coordinates turn out to be admissible, provided it holds (28) (see also [12]).

in order to obtain (C1) (as to (C2) see [12]).

We turn our attention to the regularity at the centre $\tilde{r} = 0$. Due to (28) we can extend $\rho_F$ ($U_F$, repectively) as an even (odd, respectively) $C^\infty$-function of $\tilde{r} \in [-\tilde{r}_0, \tilde{r}_0]$. This and (C2) implies that $E$ can be extended as an *even* $C^\infty([-\tilde{r}_0, \tilde{r}_0])$-function. Thus there exists $E_1 \in C^\infty([0, \tilde{r}_0^2])$ with $E_1(\tilde{r}^2) = E(\tilde{r})$. From this and $0 = E(0) - 1 = E_1(0) - 1$ it follows the existence of some $E_2 \in C^\infty([0, \tilde{r}_0^2])$ with $(E_1(\tilde{r}^2) - 1) \cdot \frac{1}{\tilde{r}^2} = E_2(\tilde{r}^2) = E_2(x_1^2 + x_2^2 + x_3^2)$. This, $E(\tilde{r}) = E_1(\tilde{r}^2)$ and (C1) imply $g_{\alpha\beta}^F \in C^\infty$ as functions of $x_\alpha$ and hence as a tensor field on $F$, which also holds for $K_{\alpha\beta}^F$ and $\rho_F$ (to be shown analogously).

This implies that the Cauchy development of the $C^\infty$-data $(g_{\alpha\beta}^F, K_{\alpha\beta}^F, \rho_F)$ is $C^\infty$ (see [12], who used [11]).